# Evidence for an axionic charge density wave in the Weyl semimetal $(TaSe_4)_2I$

J. Gooth[1]*, B. Bradlyn[2], S. Honnali[1], C. Schindler[1], N. Kumar[1], J. Noky[1], Y. Qi[1], C. Shekhar[1], Y. Sun[1], Z. Wang[3,4], B. A. Bernevig [5,6,7], C. Felser[1]

[1]*Max Planck Institute for Chemical Physics of Solids, Nöthnitzer Straße 40, 01187 Dresden, Germany.*

[2]*Department of Physics and Institute for Condensed Matter Theory, University of Illinois at Urbana-Champaign, Urbana, IL, 61801-3080, USA*

[3]*Beijing National Laboratory for Condensed Matter Physics, and Institute of Physics, Chinese Academy of Sciences, Beijing 100190, China*

[4]*University of Chinese Academy of Sciences, Beijing 100049, China*

[5]*Department of Physics, Princeton University, Princeton, New Jersey 08544, USA*

[6]*Dahlem Center for Complex Quantum Systems and Fachbereich Physik, Freie Universität Berlin, Arnimallee 14, 14195 Berlin, Germany*

[7]*Max Planck Institute of Microstructure Physics, 06120 Halle, Germany*

*johannes.gooth@cpfs.mpg.de

An axion insulator is a correlated topological phase, predicted to arise from the formation of a charge density wave in a Weyl semimetal. The accompanying sliding mode in the charge density wave phase, the phason, is an axion. It is expected to cause anomalous magneto-electric transport effects. However, this axionic charge density wave has so far eluded experimental detection. In this paper, we report the observation of a large, positive contribution to the magneto-conductance in the sliding mode of the charge density wave Weyl semimetal $(TaSe_4)_2I$ for collinear electric and magnetic fields ($E||B$). The positive contribution to the magneto-conductance originates from the anomalous axionic contribution of the chiral anomaly to the phason current, and is locked to the parallel alignment of $E$ and $B$. By rotating $B$, we show that the angular dependence of the magneto-conductance is consistent with the anomalous transport of an axionic charge density wave.

Axions refer to elementary particles that have long been known in quantum field theory,[1,2] but have yet to be observed in nature. However, it has been recently understood that axions can emerge as collective electronic excitations in certain crystals, so-called axion insulators.[3] Despite being fully gapped to single-particle excitations in the bulk and at the surface, an axion insulator is characterized by an effective action, which includes a topological $\theta \mathbf{E} \cdot \mathbf{B}$-term, where $\mathbf{E}$ and $\mathbf{B}$ are the electromagnetic fields inside the insulator, and $\theta$ plays the role of the dynamical axion field. Physically, the average value of $\theta$ is determined by the microscopic details of the band structure of the system, and gives rise to unusual magneto-electric response properties such as quantum anomalous Hall conductivities[4–9], the quantized circular photo-galvanic[4,10,11] effect, and the chiral magnetic effect.[4,12–14] The prospect of realizing an axion insulator has inspired much theoretical and experimental work. Only very recently, signatures of a dynamic axion field have been found on the surface of magnetically doped topological insulator thin films.[15–17] However, the axionic quasi-particle in these systems—the axionic polariton[3]—has so far eluded experimental detection. Alternatively, axion insulators have been predicted to arise in Weyl semimetals that are unstable towards the formation of a charge density wave (CDW).[18–23]

In their parent state, Weyl semimetals are materials in which the low-energy electronic quasiparticles behave as chiral relativistic fermions without rest mass, known as Weyl fermions.[24–27] The Weyl fermions exist at isolated crossing points of conductance and valence bands—so called Weyl nodes—and their energy can be approximated with a linear dispersion relation (Fig. 1 (a)). The Weyl nodes always occur in pairs of opposite "handedness" or chirality. At low energies and in the absence of interactions the chirality is a conserved quantum number, and the two chiral populations do not mix. Parallel electric and magnetic fields ($\mathbf{E}||\mathbf{B}$), however, enable a steady flow of quasiparticles between the left- and right-handed nodes.[28–31] This induced breaking of the chiral symmetry is a macroscopic manifestation of a quantum anomaly in relativistic field theory, giving rise to a positive longitudinal magneto-conductance in these systems. While the single-particle states in Weyl systems have been well studied experimentally, many-body interaction effects remain largely unexplored. By turning on significant interactions, a CDW can be induced in Weyl semimetals, which links the two Weyl nodes and gaps the Weyl fermions.

A CDW is the energetically preferred ground state of the strongly coupled electron-phonon system in certain quasi-one-dimensional conductors at low temperatures.[32] It is characterized by a gap in the single-particle excitation spectrum and by a gapless collective mode, formed by electron-hole pairs. The density of the electrons and the position of the lattice atoms are periodically modulated with a period larger than the original lattice constant (Fig. 1 (a)). The phase $\phi$ of this condensate is of fundamental importance for experiments: its time derivative is related to the electric current density ($j_{\text{cdw}} \sim \frac{\partial \phi}{\partial t}$) carried by the collective mode upon application of an electric field.[33] Due to its relation to the phase, this current-carrying collective mode is called the phason. In most cases, the wave vectors of the CDW are incommensurate with the lattice, and the CDW is pinned to impurities. Therefore, only upon applying a certain threshold electric field $E_{\text{th}}$ (above which the electric force overcomes the pinning forces) the CDW is depinned and free to "slide" over the lattice,[34–36] thereby contributing to the electrical conduction. The resulting conduction behavior is strongly nonlinear and the electrical resistivity drops with increasing electric field.

In a Weyl semimetal, the CDW can couple electrons and holes with different chirality, *i.e.* generates a complex mass in the Dirac equation, which mixes the two chiral populations of the system even in zero magnetic field.[18–22] The resulting gapped state is the axion insulator and the phase of the CDW is identified as the axion field ($\phi = \theta$). The phason is the axionic mode of this system and its dynamics is therefore described by the topological $\theta \mathbf{E} \cdot \mathbf{B}$-term. Compared with its high-energy version, this condensed matter realization of the axion has the advantage of being accessible in magneto-electric transport experiments.

The pairing of electrons and holes of opposite chirality and the corresponding gapping of the Weyl cones induced by the CDW implies that the chiral anomaly is irrelevant for the single-particle magneto-electric transport in the axion insulator state. We find, however, that the axionic phason mode leads to a positive longitudinal contribution to the magneto-conductance. This connection can be understood through a calculation based on a phenomenological effective field theory for the electrical conductivity (Methods). Applying $E \| B$ to the Weyl-CDW system leads to an extra anomalous contribution from the chiral anomaly to the phason current, which can be identified as a signature of the axionic character of the CDW. Aligned $E$ and $B$ fields generate an additional collective chiral flow of charges with a rate that is proportional to $E \cdot B$. In the low-field regime, the calculation yields a positive magneto-conductance contribution of $\sigma_{\mathrm{CDW},xx}(B_{//}) = c_1 + c_2 a_\chi B_{//} + c_3 (a_\chi B_{//})^2$ in the axion insulator state, where $c_1$, $c_2$ and $c_3$ are material-specific parameters, $a_\chi = \frac{1}{4\pi^2}$ is the chiral anomaly coefficient and $B_{//}$ denotes the component of $B$ that is parallel to $E$. Note that, due to threshold voltage caused by the impurity pinning in real material systems, $c_1, c_2$ and $c_3$, and therefore $\sigma_{\mathrm{CDW},xx}(B_{//})$, can in general depend on $V$ (Methods). In the low field limit, where several Landau levels are ocupied, the conductance is given by the quadratic term, recovering the characteristic, chiral anomaly-induced, positive longitudinal magneto-conductance of a non-interacting Weyl semimetal.[37–39] In particular, in this limit we obtain a $B_{//}\cos^2(\varphi)$-dependence of the magneto-conductance on the angle $\varphi$ between $E$ and $B$. In the high field limit, the magneto-conductance behaves linearly with magnetic field; similar to the ultra-quantum limit of a Weyl semimetal.[31] This is the fingerprint of the axion, allowing us to probe the presence of the axionic phason mode through its effect on the electrical transport in a Weyl-CDW condensed matter system.

Recently, a positive magneto-conductance has been observed in the CDW phase of the predicted Weyl semimetal $Y_2Ir_2O_7$,[40] because of a reduced de-pinning threshold voltage in longitudinal magnetic fields. However, the phason current was found to be independent of ***B***. To obtain evidence for the axionic phason, it is therefore desirable to go beyond these experiments.

For our study, we use the quasi-one dimensional material $(TaSe_4)_2I$ (Extended Data Fig. 1 (a)), which has recently been shown to be a Weyl semimetal[41] in its non-distorted structure (Fig.1 (b)). Its Fermi surface is entirely derived from Weyl cones with 24 pairs of Weyl nodes of opposite chirality close to the intrinsic Fermi level (within -10 meV < ($E$-$E_F$) < 15 meV). At temperatures below $T_c$ = 263 K, $(TaSe_4)_2I$ forms an incommensurate CDW phase, undergoing a Peierls-like transition,[42] which is accompanied by the opening of an approximately 260 meV gap to single particle excitations. In a recent study, the momentum (***k***)-space positions of the electronic susceptibility peaks calculated from the Fermi pockets of the Weyl points and the experimentally observed ***k*** = (222) satellite reflections in XRD measurements of the CDW phase of $(TaSe_4)_2I$ have been found to match, indicating that the CDW nests the Weyl nodes and therefore breaks the chiral symmetry.[41] In addition, $(TaSe_4)_2I$ was previously shown to exhibit non-linear electrical transport characteristics[43,44] in zero magnetic field upon crossing a threshold voltage $V_{th}$, associated with a current-carrying sliding phason. Therefore, it is a strong candidate for an axion insulator and the experimental observation of the axionic phason.

Crystals of $(TaSe_4)_2I$ grow in millimeter-long needles with aspect ratios of approximately 1:10, reflecting the one-dimensional crystalline anisotropy of the material (Methods and Extended Data Fig. 1). In total, we have measured the electrical transport of five (TaSe$_4$)$_2$I samples (A, B, C, D, E) with the electrical current $I$ applied along the $c$-axis of the crystals (Methods). On all samples, the low bias four-terminal electrical resistivity $\rho$ of the single quasi-particle excitation has been measured as a function of temperature $T$ (Extended Data Fig. 2). The current bias-dependent transport properties were then measured on samples A, B and E. All investigated samples show similar electrical transport properties. At 300 K, $\rho$ of the crystals is around $\rho$(300 K) = 1500 mΩcm (+/- 400 mΩcm) with an electron density of $n$(300 K) = (1.7 ± 0.1) × 10$^{21}$ cm$^{-3}$ and a Hall mobility of $\mu$(300 K) = (2.3± 0.1) cm$^2$V$^{-1}$s$^{-1}$ (Methods and Extended Data Fig. 3), which are in good agreement with the previous literature.[43] From further analysis of Hall measurements (Methods), we find that the Fermi level in our samples is located precisely in the middle of the CDW gap and therefore at the position of the initial Weyl cones in (TaSe$_4$)$_2$I (Extended Data Fig. 3). The data shown in the main text are obtained from measurements on sample A.

We first characterized the single-particle transport at zero magnetic field ($|\boldsymbol{B}|$ = 0 T). Fig. 1 (c) (left axis) shows $\rho/\rho$(300 K) of the single particle state on a log-scale as a function of 10$^3$K/$T$. For all samples investigated, we observe an increasing $\rho$ with decreasing $T$, consistent with the $\rho$-$T$-dependence of (TaSe$_4$)$_2$I reported in literature.[43,44] The logarithmic derivative (Fig. 1 (c), right axis) versus the inverse temperature 1/$T$ exhibits a well-pronounced peak at around the expected CDW transition temperature $T_c$ = 263 K. The saturation of the logarithmic derivative at a constant value $a_v$ above 1/$T_c$ demonstrates that $\rho/\rho$(300 K) follows the thermal activation law $\rho/\rho$(300 K) ~ exp[$\Delta E/k_B T$], allowing us to deduce the size of the single particle gap $\Delta E$ in units of $k_B$. Averaging over the saturated range, we obtain a value of $a_v$ = 0.72 ± 0.04 and thus $\Delta E$ =(259 ± 14) meV, which is in excellent agreement with literature.[42,43]

Next, we investigate the motion of the CDW at zero magnetic field in the pinning potential created by impurities. Fig. 2 (a-c) shows the variation of the measured voltage $V$ as a function of the applied DC current $I$ at four selected temperatures. Consistent with a sliding phason mode, non-linearity in the $V$-$I$ curves appears below $T_c$ at high bias currents i.e. high voltages (Extended Data Fig. 4). Accordingly, the $V$-$I$ curves can be well represented by a simple phenomenological model, based on Bardeen's tunneling theory, as suggested by previous electrical transport experiments on CDWs[45] (Methods, Fig. 2 (d) and Extended Data Fig. 5). To gain further insights into the collective charge transport mechanism, we calculate the differential electrical resistance d$V$/d$I$ from the $V$-$I$ curves and plot it versus the corresponding $V$ for various $T$ (Fig. 2 (e)). For all temperatures investigated, the d$V$/d$I$ is constant at low $V$. However, below $T_c$, the d$V$/d$I$ decreases above a threshold voltage $V_{th}$, indicating the onset of the collective phason current.[32] $V_{th}$ is determined as the onset of the deviation from the zero-baseline of the second derivatives d$V^2$/d$I^2$ (see exemplarily Extended Data Fig. 6). The magnitude of the corresponding $E_{th} = V_{th} / L$ ($L$ is the distance between the voltage probes) and its $T$-dependence (Fig. 2 (f)) are in agreement with literature.[43] Further analysis suggests that the non-linear $V$-$I$ characteristics are not the result of local Joule heating nor a consequence of contact effects (Methods and Extended Data Fig. 7). These observations provide evidence for a propagating, current-carrying CDW state.

We now turn to testing the axionic nature of the phason. For this purpose, we measure the non-linear $V$-$I$ characteristics at fixed $T$, but now in the presence of a background magnetic field. We apply a magnetic field between -9 and +9 T in 1 T steps, oriented perpendicular and in parallel to the direction of current flow. All experimental observations are symmetric in ***B*** (compare Fig. 3 and Extended Data Fig. 8). As shown in Fig. 3 (a) and (b), no ***B***-modulation of the $V$-$I$ characteristics and d$V$/d$I$ can be seen within the measurement error of 0.1% for ***B*** perpendicular to $I$ ($E\perp B$) across the whole magnetic field and temperature range investigated.

To enable a direct comparison to the theory, we calculate the differential magneto-conductance $\Delta dI/dV(\bm{B}) = dI/dV(\bm{B}) - dI/dV\ (\bm{B} = 0\ \text{T})$. No $\Delta dI/dV(\bm{B})$ is observed (Fig. 3 (c) and (d)) in the perpendicular field configuration. However, when the magnetic field is aligned with $I$ ($\bm{E}||\bm{B}$) (Fig. 3 (e) and (f)), the $V$-$I$ characteristics and the $dV/dI$ display a strong dependence on the magnitude of the magnetic field $|B_{//}|$ above $V_{\text{th}}$ and at low temperatures. Comparing the electrical currents at fixed $T$, $B_{//}$ and $V$, we find a strong enhancement of $I$ with increasing $|B_{//}|$. Because $V_{\text{th}}$ itself is independent of $|B_{//}|$ in any direction up to ±9 T (Fig. 3 (b) and (f)), the observed $\bm{B}$-dependence seems to originate from a field-dependent contribution to the phason current. As shown in Fig. 3 (g) and (h), we consistently observe a large positive $\Delta dI/dV(\bm{B})$ for $\bm{E}\|\bm{B}$ and $T \leq 155$ K (see Extended Data Fig. 9 and 10). The profile of which can be well described by a quadratic function below 7.5 T, according to the prediction for an axionic phason. To demonstrate the quadratic-behavior at low magnetic fields, we perform second-order polynomial fits $d_1 B_{//} + d_2 B_{//}^{\ 2}$ to the experimental data, with the open parameters $d_1$ and $d_2$, and find that $d_1 B_{//} << d_2 B_{//}^{\ 2}$ (Extended Data Fig. 11). By comparison to our theoretical model for transport (Methods), the dominant quadratic behavior for all $B_{//} < 7.5$ T investigated is consistent with the estimated Fermi level in our samples near the position of the initial (undistorted) Weyl nodes. Above 7.5 T, $\Delta dI/dV(\bm{B})$ deviates from the quadratic function, which may be understood by a transition to the quantum limit, or a $B$-dependent axial charge relaxation time at high magnetic fields.[39] We note that $\Delta dI/dV(\bm{B})$ in our experiment depends on $V$ (Extended Data Fig. 11). As explained above, we attribute this observation to the impurity pinning of the CDW in our sample. The associated pinning potential induced strongly non-ohmic $I$-$V$ characteristics, which are not included in our theoretical calculation.

To test the angular variation of the positive $\Delta dI/dV(\bm{B})$, we perform $V$-$I$ measurements at 80 K and fixed 9 T for various angles $\varphi$ (Fig. 4 (a)). In accordance with the theoretical predictions for the axionic phason current, we find that the $\Delta dI/dV(\varphi)$ is described by a $\cos^2(\varphi)$- dependence. The observed pattern of angular dependence and the consistent functional dependence of the longitudinal $\Delta dI/dV(B)$ on $B$ are the fundamental signatures of the axionic phason and the associated chiral anomaly. This supports the identification of $(TaSe_4)_2I$ as an axion-insulator.

In conclusion, our measurements reveal a positive longitudinal magneto-conductance in the sliding mode of the quasi-one-dimensional CDW-Weyl semimetal $(TaSe_4)_2I$, a signature that is linked to the presence of an axionic phason. In short, we theoretically and experimentally find a $B_{//}^{-2}$-dependence of the longitudinal magneto-conductance and a $\cos^2$-dependence on the relative orientation between $I$ and $B$. Our results show that it is possible to find experimental evidence for axions, particularly elusive in other contexts, in strongly correlated topological condensed matter systems.

**FIGURES**

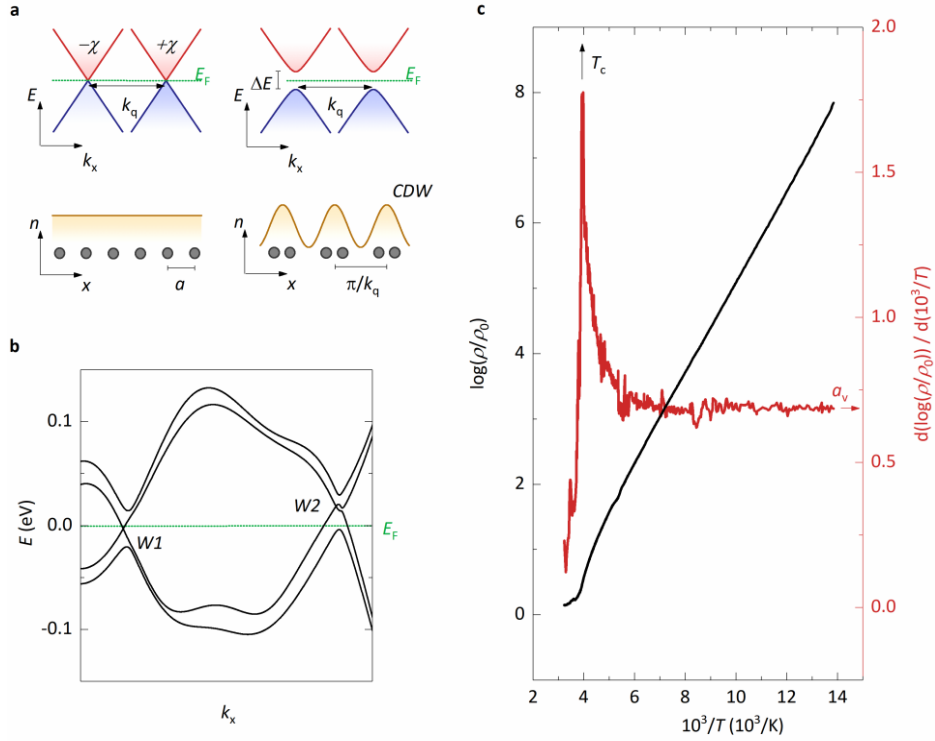

**Figure 1 | Charge density wave (CDW) in the Weyl semimetal (TaSe$_4$)$_2$I. a,** Schematics of the CDW formation in a simple Weyl semimetal. Upon the periodic modulation of the carrier density $n$ in real space $x$, the two Weyl cones of opposite chirality $+\chi$ and $-\chi$ in energy ($E$)-momentum($k_x$) space become gapped at the Fermi energy $E_F$. Accordingly, the original lattice spacing $a$ is modulated by $\pi/k_q$, where $k_q$ is the initial distance of the Weyl nodes in $k_x$. Note that in (TaSe$_4$)$_2$I, there exist multiples of 2 Weyl points at $E_F$ and therefore the CDW formation might be more complex in this material. **b,** Ab-initio calculations of the band structure reveal two types of Weyl nodes (*W*1 and *W*2) near $E_F$ along $z$. **c,** The electrical resistivity $\rho$, normalized to the 300K-value $\rho_0$, (left axis, black) and its logarithmic derivative (right axis, red) as a function of $10^3/T$, where $T$ is the temperature in Kelvin. The derivative peaks at the CDW transition temperature $T_C$ = 263 K, which allows us to extract of the single-particle gap $\Delta E$ from the average value of the logarithmic derivative $a_v$ above $10^3/T = 7·10^3$ K$^{-1}$.

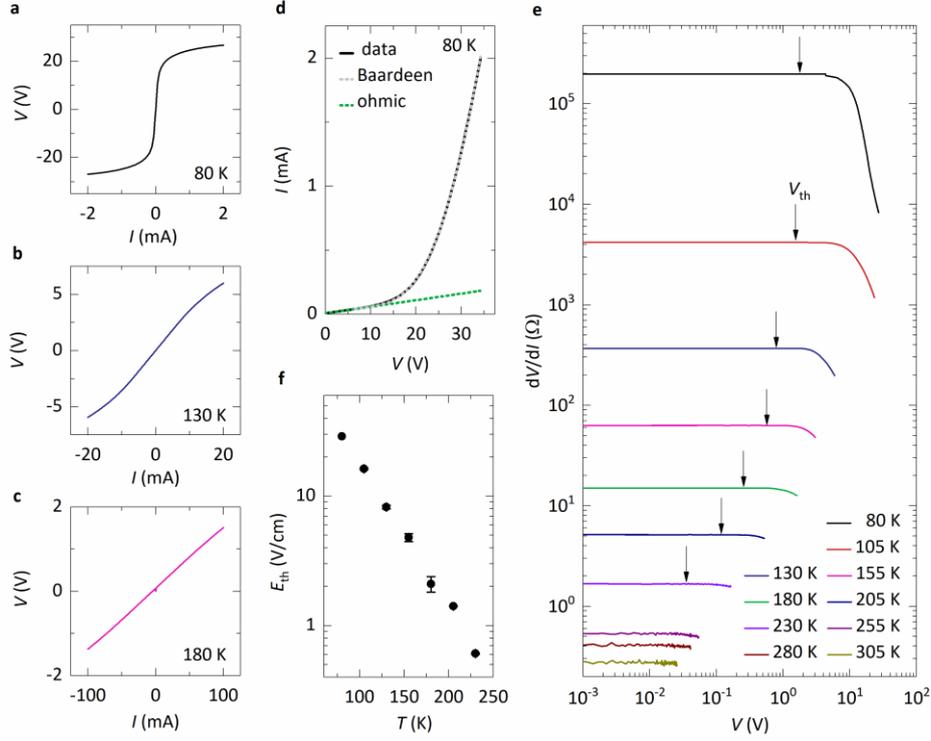

**Figure 2 | Collective propagation of the charge density wave in (TaSe$_4$)$_2$I. a,** Voltage($V$)-current($I$) characteristics measured across a $V$-probe distance $L$ of 2 mm at representative temperatures of 80 K, **b**, 130 K and **c**, 180 K without a magnetic field applied. **b,** $I$ versus $V$ at 80 K. The light dotted lines are fits to Bardeen's tunneling theory of sliding CDWs for V > $V_{th}$. The green dotted lines denote linear fits for V < $V_{th}$, corresponding to the ohmic current contribution. **f**, Threshold electric field $E_{th} = V_{th}/L$ as a function of $T$ calculated from the $T$-dependent $V_{th}$ (Extended Data Fig. 5(i)). The error bars denote the variation between $+V_{th}$ and $-V_{th}$ due to hysteresis effects. **e,** Differential resistance d$V$/d$I$ as a function of $V$ at various temperatures $T$. The arrows point at the threshold voltage $V_{th}$, which is defined as the voltage at which the second derivative d$V^2$/d$I^2$ becomes non-zero and decreases with increasing $V$ (Extended Data Fig. 6). The decrease in d$V$/d$I$ is the signature of the collective motion of the CDW.

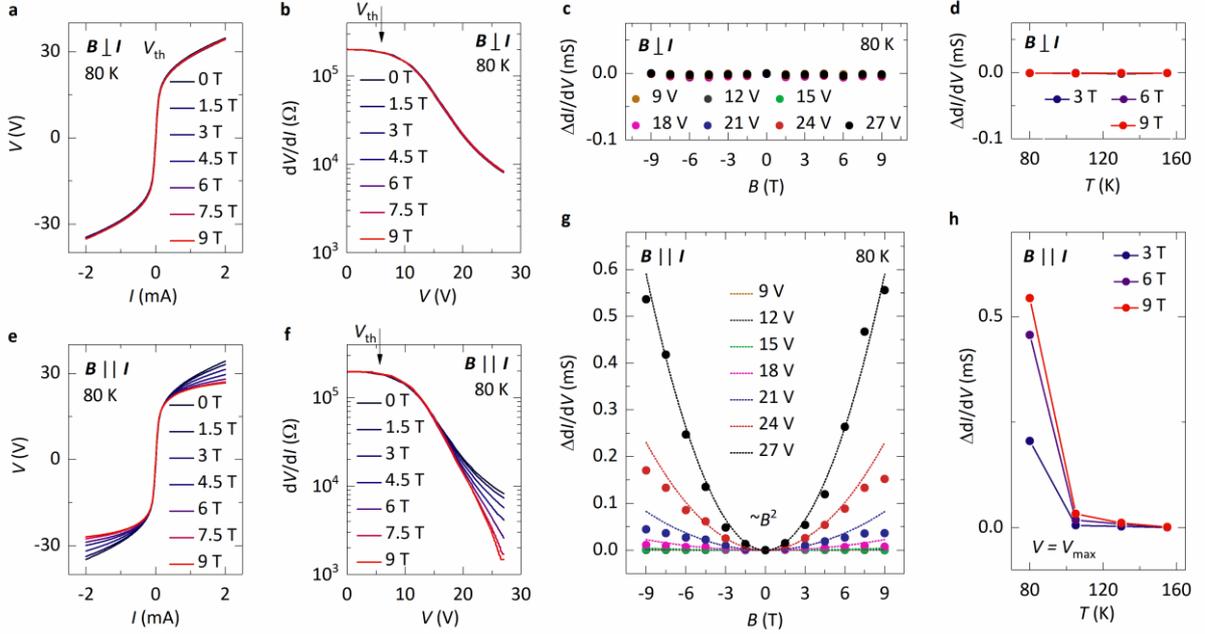

**Figure 3 | Evidence for an axionic phason in (TaSe₄)₂I.** The transport measurements are carried out in a magnetic field *B*. **a**, *V-I* characteristic and **b**, d*V*/d*I*, at 80 K at various values of the magnetic field *B,* applied perpendicular to the current direction (*I*⊥*B*). **c**, *B*-dependent variation of the differential conductance Δd*I*/d*V*(*B*) = d*I*/d*V* (*B*) - d*V*/d*I* (0 T) at 80 K and at various *V* and **d,** *T*-dependence of the Δd*I*/d*V*(*B*). The electrical transport does not show a discernible dependence on the perpendicular field. However, for magnetic fields *B* parallel to *I* (*B*||*E*), **e**, the current at voltages above $V_{th}$ is significantly enhanced with increasing *B* and **f**, the differential resistance d*V*/d*I* is strongly suppressed. **g**, The longitudinal magneto-conductance Δd*I*/d*V*(*B*) becomes increasingly positive with increasing *V* and is well described by quadratic fits at low magnetic fields (the fit parameters are shown in Extended Data Fig. 11). **h**, The positive longitudinal Δd*I*/d*V*(*B*) diminishes at higher *T,* when plotted at the highest applied voltage $V_{max}$ at each specific *T*: $V_{max}$(80 K) = 27 V, $V_{max}$(105 K) = 24 V, $V_{max}$(130 K) = 6 V), $V_{max}$(155 K) = 3 V.

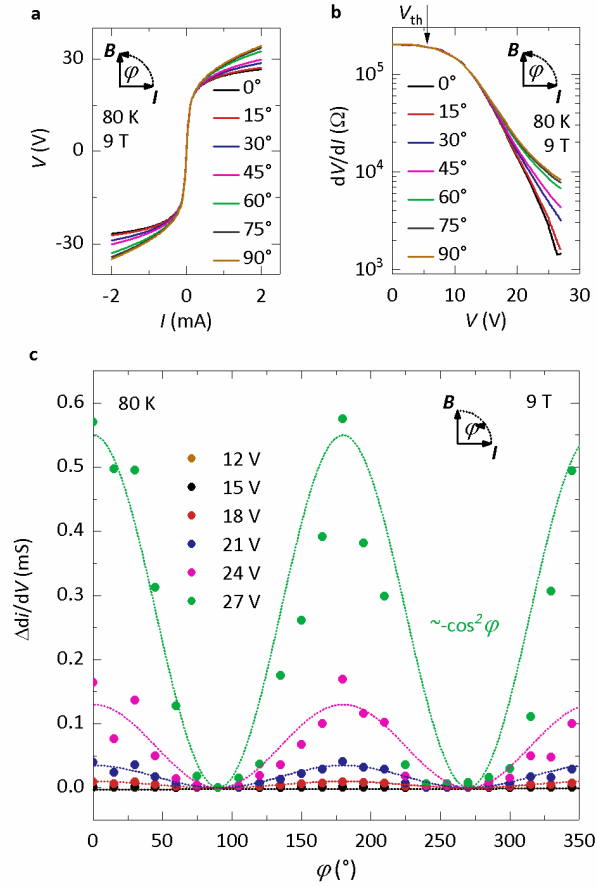

**Figure 4 | Angular dependence of the axial current. a**, The angular dependence is inferred from measurements of the *V-I* characteristics in a tilted $B(\varphi)$ of 9 T at 80 K, where $\varphi$ denotes the angle with respect to the applied current *I*. **b**, The dependence of d*V*/d*I* on the angle $\varphi$ for voltages above $V_{th}$. **c**, $\Delta dI/dV$ at 9 T is plotted against $\varphi$ for various *V*. The measured data is reasonably well-represented by a fits to $\cos^2(\varphi)$. The axial current peaks when $\varphi \rightarrow 0°$ and 180°.

**METHODS**

**Phenomenological theory of the MR in an axionic CDW system**

In this work, we model the magneto-transport in a CDW-Weyl system (i.e. axion insulator)[19] using an effective field theory. For simplicity, we start with a quasi-one-dimensional Weyl semimetal with exactly two Weyl nodes. The realistic situation of $(TaSe_4)_2I$ is much more complicated, but we here restrict to this simpler case. We assume that the Fermi surface consists of disconnected pockets surrounding these two Weyl nodes. Furthermore, we assume that the CDW ordering wave vector is commensurate with the Weyl node separation $\boldsymbol{k_q}$ and use natural units. To simplify the discussion, we will neglect the effects of impurity pinning. We will then discuss the effects of impurity pinning by examining the modifications to the *B=0* conductivity.

**Effective field theory for the semi-classical limit, where many Landau levels are occupied.**

Our strategy is to introduce an order parameter

$$|\Delta|e^{i\phi} = \langle c^{\dagger}_{-\frac{k_q}{2}+\delta k} c_{\frac{k_q}{2}+\delta k}\rangle \quad (1)$$

for the CDW order by decoupling the electron-phonon interaction Hamiltonian. Here $c_{\frac{k_q}{2}+\delta k}$ is an operator which annihilates an electron displaced by wavevector $\delta \boldsymbol{k}$ from the Weyl node centered at $\pm\frac{k_q}{2}$. Integrating out the electrons and neglecting the fluctuations in the gapped amplitude mode $|\Delta|$, we then arrive at an effective field theory for the phason mode $\phi$ of the CDW. Let us confine ourselves to zero temperature. Neglecting the chiral charge of the Weyl nodes (we will add it later), we then find from Anderson[46] the action (working in units where $\hbar = 1$)

$$S_0 = \frac{m_F}{2} N(0) \int d^3x dt \left[ -\frac{1}{2}\left(\frac{\partial \phi}{\partial t}\right)^2 + \frac{1}{2} c_\perp (\widehat{k_q} \times \nabla \phi)^2 + \frac{1}{2} c (\widehat{k_q} \cdot \nabla \phi)^2 - 2e^* \frac{v_F}{m_F} \widehat{k_q} \cdot E \phi \right].$$

(2)

Here we have defined

$$m_F = 4 \frac{|\Delta|^2}{\lambda \omega_{k_q}^2} \quad (3)$$

with $\lambda$ the electron-phonon coupling, and $\omega_{k_q}$ the phonon energy at wavevector $k_q$. Furthermore, $N(0)$ is the density of states at the Fermi level, $v_F$ is the Fermi velocity in the direction of the CDW wave vector, $\widehat{k_q}$, and $e^*$ is the screened electron charge. The velocities $c_\perp$ and $c$ correspond respectively to the transverse and longitudinal (phason) velocities, respectively. The final term in (2) corresponds to the sliding (Froelich) mode, excited by an electric field aligned with the CDW ordering vector.

Additionally, we find from the chiral anomaly the contribution[1,8]

$$S_c = \frac{e^2}{4\pi^2} N(0) \int d^3x dt \, \phi E \cdot B \quad (4)$$

This contribution to the action emerges from the chiral anomaly in the undistorted Weyl semimetal; as shown in Ref. 1, the formation of the CDW causes the amplitude of the chiral anomaly to depend on the phason field. Let us now consider the effect of a uniform external magnetic field. Since we are interested in response to uniform fields, we will restrict ourselves to spatially homogeneous configurations $\phi(x, t) = \phi(t)$. By varying the action $S = S_0 + S_c$ with respect to $\phi$, we find, from the equations of motion, that $\phi$ satisfies

$$\frac{\partial S}{\partial \phi} = \frac{N(0) m_F}{4} \frac{\partial^2 \phi}{\partial^2 t} - E \cdot \left( N(0) v_F e^* \widehat{k_q} - \frac{e^2}{4\pi^2} B \right) \quad (5)$$

To include the effects of damping, we follow Rice et al.[47] and add a phenomenological damping term to this equation. Since we are interested in the dynamics of the sliding mode above the de-pinning transition, we will not consider a phenomenological pinning term. Defining an effective damping rate $\Gamma$, we thus find

$$\frac{N(0)m_F}{4}\frac{\partial^2\phi}{\partial^2 t} + \Gamma\frac{\partial\phi}{\partial t} = \boldsymbol{E}\cdot\left(N(0)v_F e^*\widehat{\boldsymbol{k}_{\mathrm{q}}} - \frac{e^2}{4\pi^2}\boldsymbol{B}\right) \quad (6)$$

We can solve this equation in the long-time limit to find the steady-state solution

$$\phi(t) = \phi_0 + \frac{t}{\Gamma}\boldsymbol{E}\cdot\left(N(0)v_F e^*\widehat{\boldsymbol{k}_{\mathrm{q}}} - \frac{e^2}{4\pi^2}\boldsymbol{B}\right) \quad (7)$$

Next, we look at the current density

$$\boldsymbol{j} = \frac{\delta S}{\delta \boldsymbol{A}} \quad (8)$$

carried by the CDW sliding mode, where $\boldsymbol{A}$ is the electromagnetic vector potential. Varying the action and restricting to position-independent configurations of $\phi$, we find

$$\boldsymbol{j} = \left(N(0)v_F e^*\boldsymbol{k}_{\mathrm{q}} - \frac{e^2}{4\pi^2}\boldsymbol{B}\right)\frac{\partial\phi}{\partial t} \quad (9)$$

Finally, combining Eq. 7 with Eq. 9, we find that

$$j^i = \sigma_{ij}(\boldsymbol{B})E_j \quad (10)$$

with the conductivity tensor given by

$$\sigma_{ij}(\boldsymbol{B}) = \frac{1}{\Gamma}\left(N(0)v_F e^*\widehat{\boldsymbol{k}_{\mathrm{q}}} - \frac{e^2}{4\pi^2}\boldsymbol{B}\right)_i\left(N(0)v_F e^*\widehat{\boldsymbol{k}_{\mathrm{q}}} - \frac{e^2}{4\pi^2}\boldsymbol{B}\right)_j \quad (11)$$

In particular, the magneto-conductivity is given by

$$\frac{\sigma_{ij}(\boldsymbol{B}) - \sigma_{ij}(0)}{\sigma_{ij}(0)} = -\frac{e^2}{4\pi^2 N(0)v_F e^*}\left(\widehat{k_{\mathrm{q},i}}B_j + B_i\widehat{k_{\mathrm{q},j}}\right) + \frac{e^2}{4\pi^2(N(0)v_F e^*)^2}B_i B_j. \quad (12)$$

Note that there are two types of contributions in Eq. 12. The first term depends on the relative orientation of the electric and magnetic fields relative to the CDW wavevector, and originates from a mixing between the chiral anomaly ($\boldsymbol{B}$-dependent) and conventional ($\boldsymbol{B}$-independent) contributions to the CDW dynamics in 7 and 9. The quadratic in *B* contribution to the magneto-conductivity, on the other hand, has a purely chiral-anomaly origin. Although it corresponds to current carried by the sliding mode, its origin is entirely due to the underlying Weyl semi-metal. Hence, like in the Weyl semimetal, this contribution to the magneto-conductivity is independent of the Weyl separation (i.e. CDW ordering) vector.

Now note that in the idealized case when the unordered phase is a perfect Weyl semimetal $N(0) \rightarrow 0$. In this case, the second term in Eq. 12 dominates over the first, and we recover the characteristic quadratic positive magneto-conductivity of a Weyl material. In particular, in this limit we also recover the characteristic $\cos^2$- dependence of the magnetoresistance on the angle between *E* and *B*. More generally, the quadratic term in Eq. 12 will dominate when

$$N(0) v_F e^* \ll e^2 |\boldsymbol{B}|, \quad (14)$$

Hence, we expect a positive quadratic magneto-conductivity in this regime.

**Microscopic argument for the quantum limit, where only the zeroth Landau level is occupied.** To gain some intuition for how the sliding mode can influence the magnetoresistance in the quantum limit, let us take our Weyl-CDW, and apply a large magnetic field along the $\widehat{\boldsymbol{k}_q}$ direction. We take the magnitude of *B* to be large enough that only the zeroth Landau level is occupied when electron-phonon interactions are neglected. In this case, we have a truly quasi-one-dimensional system, with conserved momentum $Q = \boldsymbol{k} \cdot \widehat{\boldsymbol{k}_q}$; each state is largely degenerate, with degeneracy given by the total flux $|\boldsymbol{B}|A/2\pi$ through the system (measured in units of the flux quantum, with *A* the cross-sectional area). Let us now reintroduce the electron-phonon interaction; this causes the formation of a CDW ground state. Because the Fermi points of the Weyl nodes disperse entirely in the lowest Landau level, the low energy physics of this transition matches that of a true one-dimensional CDW transition, provided we multiply all results by the Landau level degeneracy $|\boldsymbol{B}|A/2\pi$. In particular, Lee, Rice, and Anderson showed[46] that the sliding mode of the CDW contributes $e^2n/m^*\Gamma$ per channel to the electrical conductivity in the $\widehat{\boldsymbol{k_q}}$-direction, where $m^* = m(1 + m_F)$ is the electron effective mass, $n$ is the one-dimensional charge carrier density, and $\Gamma$ is the phenomenological damping rate (phonon lifetime). Multiplying this by the number of channels we find in the quantum limit and for current and electric field aligned with the magnetic field (and therefore CDW) direction that

$$\sigma_{ij}(\boldsymbol{B})\widehat{\boldsymbol{B}}_i\widehat{\boldsymbol{B}}_j \to \frac{e^2n}{m^*\Gamma}|\boldsymbol{B}|\ , \quad (15)$$

analogous to the magnetoresistance in the ultra-quantum limit of a Weyl semimetal.[38]

**Comment on real material systems**

Having justified the origin of the negative magnetoresistance in a simplified model, we now discuss how pinning effects will change the phenomenology. First, since CDWs are always pinned by impurities in real material systems, sliding only occurs upon reaching a certain threshold electric field, i.e. threshold voltage $V_{th}$. In addition, transport measurements always also involve single charge carriers, thermally exited above the Peierls gap $\Delta E$, which carry electrical current in parallel to the sliding CDW mode. Therefore, the *B=0* electrical conductivity, i.e. electrical resistivity, will in general be non-ohmic and depend on the applied bias voltage *V*. Below the Peierls transition, the total electrical conductivity for a CDW system in zero magnetic field ( $\sigma_{ij,\text{total}}$ ) is given by is given Bardeen's tunneling theory[45]:

$$\sigma_{ij,\text{total}} = \sigma_{\text{SP}} + \sigma_{\text{CDW},0}\left(1 - \frac{V_T}{V}\right) \cdot \exp\left(-\frac{V_0}{V}\right) \quad (16)$$

Here, $\sigma_{\text{SP}}$ denotes the ohmic conductivity contribution of the thermally exited single charge carriers, $V_T$ is the threshold voltage and $\sigma_{CDW,0}$ and $V_0$ are *V*-independent parameters that usually depend on the temperature *T*. This is derived by assuming that pinning by impurities creates a gap $E_g < \Delta E$ to CDW sliding. Eq. 16 then follows from considering Zener tunneling of the CDW across the gap. $V_0$ is closely related to the dynamics of the sliding CDW and can be expressed as $V_0 = E_g/2\xi e^*$, where $\xi$ is the coherence length, analogous to superconductors. Because $E_g$ and $e^*$ do not depend on $T$, the temperature-dependence of $V_0$ provides a measure for the temperature-dependence of $\xi$, which generally decreases with increasing $T$. $\sigma_{CDW,0}$ is a measure of the number of carriers forming the CDW and has been observed to increase with increasing $T$. The validity of Eq. 16 for CDW systems has been confirmed by many experiments, fitting the measured $I$-$V$ characteristics to the current $I_{\text{total}}(V) = I_{\text{SP}}(V) + I_{\text{CDW}}(V)$, corresponding to Eq. 16 with $I_{\text{SP}}(V) = G_{\text{sp}}V$ accounting for the single-particle contribution and $I_{\text{CDW}}(V) = I_0(V-V_T)\exp(-V_0/V)$ for the sliding CDW. $G_{\text{sp}}$ (the single particle electrical conductance), $I_0$ (corresponding to $\sigma_{CDW,0}$) and $V_0$ are the open fit-parameters. Importantly, Fleming and Grime showed that there is no added conductivity below $V_{\text{th}}$ and thence $I_{\text{total}}(V) = I_{\text{SP}}(V)$ for $V < V_{\text{th}}$. $V_{\text{th}}$ can be determined independently as shown in the Extended Data Fig.. 6. Fitting $I_{\text{total}}(V)$ to our measurement data at zero-magnetic field for $V > V_{\text{th}}$ (Extended Data Fig. 5), we find excellent agreement with Bardeen's tunneling theory. The extracted $V_0$ decreases and $I_0$ increases with increasing temperature, in agreement with previous reports for other CDW systems. Fitting our measured $I$-$V$ characteristics to the corresponding current-expression $I_{\text{total}}(V) = I_{\text{SP}}(V) + I_{\text{CDW}}(V)$, with $I_{\text{SP}}(V) = G_{\text{sp}}V$ and $I_{\text{CDW}}(V) = I_0(V-V_T)\exp(-V_0/V)$, using the single particle electrical conductance $G_{\text{sp}}$ and $I_0$ and $V_0$ as the open fit-parameters, we find excellent agreement with Bardeen's tunneling theory (Fig. 2 (d) and Extended Data Fig. 5).

The impurity pinning in real materials, such as in our samples, has important consequences on the experimental detection of the axionic response of CDWs. As mentioned above, the zero field conductivity tensor will develop a nonlinear dependence on $V$. Additionally, the single-particle current will always be measured in parallel to the sliding CDW in the experiments. Putting these facts together, we see from Eq. 12 that the measured magneto-conductance $\frac{\Delta I}{\Delta V(B)} \propto \sigma_{xx}(B) - \sigma_{xx}(0)$ will become non-ohmic and will in general depend on *V*. Hence, also the amplitude of the characteristic $\cos^2$- dependence of the magneto-conductivity on the angle between ***E*** and ***B*** will become *V*-dependent.

**(TaSe$_4$)$_2$I crystal structure and single-crystal growth**

(TaSe$_4$)$_2$I is a quasi-one dimensional material with a body-centered tetragonal lattice (Fig. 1 (b)). The Ta-atoms are surrounded by Se$_4$ rectangles, and form chains aligned along the *c*-axis. These chains are separated by $I^-$ ions. Single crystals of Ta$_2$Se$_8$I were obtained via a chemical vapor transport method using Ta, Se and I as starting materials.[48] A mixture with composition Ta$_2$Se$_8$I was prepared and sealed in an evacuated quartz tube. The ampoule was inserted into a furnace with a temperature gradient of 500 to 400°C with the educts in the hot zone. After two weeks, needle-like crystals had grown in the cold zone (Extended Data Fig._1). Energy-dispersive X-ray spectroscopy reveals a Ta:Se:I-ratio of 17.75:72.95:9.3, which is matches the theoretical ratio of 18.18:72.73:9.09 within the measurement precision of 0.5.

**Electrical Transport Measurements**

All electrical transport measurements were performed in a temperature-variable cryostat (Quantum Design), equipped with a 9 T magnet. To avoid contact resistance effects, only four-terminal measurements were carried out. In order to reduce any possible heating effects during current-sweeps, the samples were covered in highly heat-conducting, but electrically insulating epoxy.[43] The temperature-dependent electrical resistivity curves were measured with standard low-frequency (*f* = 6 Hz) lock-in technique (Keithley SR 830), applying a current of 100 nA across a 100 MΩ shunt resistor. The applied bias current *I* = 100 nA for these measurements was chosen such that the corresponding electric field *E* = *IR*/*L* does not exceed the threshold field $E_{th} > E$, where $L$ is the distance between the voltage probes. The determination of $E_{th}$ will be explained later. The current bias-dependent transport properties were then measured on samples A, B and E, because their closely adjacent electrical contacts of $L < 3$ mm allowed for crossing the threshold field $E_{th}$ with experimentally accessible currents ($IR < 60$ V) in our setup. The current bias-dependent transport experiments were limited to temperature $T \geq 80$ K. This is because $E_{th}$ exponentially increases with decreasing $T$. Given the minimal contact resolution of $> 0.5$ mm in our bulk samples, no switching could be achieved in our samples below 80K, because the corresponding threshold voltages exceed the measurable voltage range ($< 60$ V). The voltage-current characteristics were measured DC with a Keithley 6220 current source ($10^{14}$ Ω output impedance) and a Keithley 2182A/E multimeter.

**First principle calculations**

For the ab-initio investigations we employ density-functional calculations (DFT) as implemented in the VASP package.[49] We use a plane wave basis set and the generalized-gradient approximation (GGA) as a description for the exchange-correlation potential.[50] As a second step we create Wannier functions from the DFT states using Wannier90[51] and extract the parameters for a tight-binding Hamiltonian to further analyze the band structure.

**Estimate of the Fermi level position from the Hall measurements.**

In order to estimate the Fermi level position in our samples, we have performed Hall measurements on sample D. Because the single-particle resistivity of all our samples is similar throughout the whole temperature-range, the Hall measurements is representative for the whole set of samples presented in this work. The device for the Hall measurements is shown in Extended Data Fig. 3 (a). Contacts 1 and 4 are used for electrical current injection, contacts 2 and 3 are used to probe the longitudinal voltage along the sample and contacts 5 and 6 are used to measure the Hall voltage $V_{5,6}$ across the sample. The magnetic field $B$ is applied perpendicular to the measurement plane. The Hall resistance as a function of magnetic field $B$ is then calculated using $R_H = V_{5,6}/I$ at fixed temperatures $T$. We have chosen the electrical current $I$ applied at each temperature small enough to only probe the single-particle transport, but high enough to resolve a Hall signal $V_{5,6}$. In particular, we have used $I = 1$ μA at 125 K, $I = 10$ μA at 155 K and 185 K, $I = 100$ μA at 215 K and 245 K, as well as $I = 2$ mA at 300 K. Although the data is quite noisy, we observe a strong $B$-dependence of $R_H$ (Extended Data Fig. 3 (c)-(h)). Below 125 K, the longitudinal resistance of the sample became too high to measure the Hall response. As reported previously,[44] the Hall resistance for the low temperatures is linear in $B$. At 300, however, $R_H$ becomes non-linear in $B$ at high fields, which is typical for semiconductors and insulators at elevated $T$, because both, electrons and holes become thermally activated across the gap. At low temperatures, the band closer to the Fermi level dominate. As discussed previously, even in such a one-dimensional material as (TaSe$_4$)$_2$I, the Hall effect is well defined, because to the first order in $B$, the Lorentz force does not depend on the transverse effective mass.[44] Recalling the standard expression for the low field Hall conductivity (single charge carrier approximation), we estimate the carrier concentration $n = (dR_H/dB)^{-1}(ed)^{-1}$ as function of $T$ from the slope $dR_H/dB$ of the linear fits to the magnetic field-dependent $R_H$. e is the elementary charge and $d = 300$ mm the height of sample D. As seen in (Fig. 3 (i)), $n$ decreases by orders of magnitude when decreasing $T$ from 300 K to 125 K, consistent with the increasing longitudinal single particle resistivity (Extended Data Fig. 3 (b)). At the CDW transition temperature $T_C$ a jump in n is observed, consistent with the opening of a single-particle gap. At 300 K, the estimated carrier concentration is $n = (1.7 \pm 0.1) \times 10^{21}$ cm$^{-3}$ leading to a mobility of $\mu = (2.3 \pm 0.1)$ cm$^2$V$^{-1}$s$^{-1}$, using the Drude model $\mu = (en\rho)^{-1}$ with the electrical resistivity of $\rho$

= 1562 μΩcm at 300 K. Both, $n$ and $\mu$ are in good agreement with the values previously reported for $(TaSe_4)_2I$.[44]

Using the effective mass $m^* = 0.4\ m_0$ ($m_0$ is the free electron mass) obtained from ARPES measurements for $T < T_C$[43] and a single band model, we subsequently determine the Fermi level position $E_F$ in our sample with respect to the conduction band edge:

For a band gap material, the carrier concentration is given by

$$n = N_c \cdot \exp(\frac{E_F}{k_B T}), \quad (17)$$

where $N_c = 2\left(\frac{2\pi m^* k_B T}{h^2}\right)^{\frac{3}{2}}$, $k_B$ is the Boltzmann constant and $h$ denotes the Planck constant. Transforming equation 17, we obtain the description for $E_F$ that only depends on $T$, $m^*$ and $n$:

$$E_F = ln\left(\frac{1}{2}n\left(\frac{2\pi m^* k_B T}{h^2}\right)^{-\frac{3}{2}}\right)k_B T, \quad (18)$$

The result of which as a function of $T$ is shown in Extended Data Fig. 3 (j). Within the error bar, the estimated $E_F$ is independent of temperature for $T < T_C$, giving a mean value of $E_F = (129 \pm 10)$ meV below the conduction band edge of the CDW gap. Comparing this number to the experimentally obtained CDW gap size of $\Delta E = (159 \pm 5)$ meV, we find that the Fermi level in our $(TaSe_4)_2I$ samples is precisely in the middle of the CDW gap ($E_F = ½\ \Delta E$). As shown previously *via* DFT calculations and ARPES measurements, the CDW gap opens symmetrically around the crossing points of the initial electronic band structure.[42] Therefore, the Fermi level position is in the center of the initial Weyl cones.

**Excluding other intrinsic and extrinsic origins for the non-linear VI-curves.**

The following analysis suggests that the non-linear $V$-$I$ characteristics are not the result of local Joule heating nor a consequence of contact effects, but rather originate from an electrically driven intrinsic process in the $(TaSe_4)_2I$ crystals, consistent with a sliding CDW. First, we exclude thermally driven switching due to Joule heating. Thermally driven switching corresponds to a local rise of the temperature in the $(TaSe_4)_2I$ sample above $T_c$. Such a temperature-raise would be determined by the power dissipated in the sample and the electronic circuitry. In the worst-case scenario, all of the $V$-$I$ Joule heating power is dissipated within the $(TaSe_4)_2I$ crystals. Hence, the necessary dissipated power for the switching would approach zero as $T \rightarrow T_c$ from below. However, the electrical power $P_{th} = V_{th}^2/[dV/dI(V_{th})]$ obtained at $V_{th}$ from our experiments (Extended Data Fig. 7 (a)) does not approach zero as $T \rightarrow T_c$. Instead, $P_{th}$ decreases with decreasing $T$. This is precisely the opposite of what we would expect from thermally driven switching. Secondly, we rule out contact effects as the cause of the non-linear threshold behavior in our experiments. Therefore, we exemplarily show $V_{th}$ as a function of the distance of the voltage probes $L$ at 80 K in Extended Data Fig. 7 (b) for $L$ = 1 mm, 2 mm and 3 mm. The observed linear $V_{th}$–$L$-dependence implies that the transition is driven by the electric field and not by the absolute magnitude of $I$, which justifies our definition of $E_{th} = V_{th} / L$. It also demonstrates that the transition originates from a bulk effect, because the contacts in all of our devices are of identical size and, hence, switching in the contacts would have no length-dependence. Therefore, any change in switching properties must result from the $(TaSe_4)_2I$ crystal itself.

**Excluding other intrinsic and extrinsic origins for the positive differential magneto-conductance.**

Here, we address the concern that the experimentally detected positive differential magneto-conductance $\Delta dI/dV(B)$ in $(TaSe_4)_2I$ could arise from an alternative origin than the axionic nature of the CDW. Several intrinsic and extrinsic origins of the positive $\Delta dI/dV(B)$ in $(TaSe_4)_2I$ can be excluded from the experimental observations and analysis: (i) Inhomogeneous current distribution caused by quenched disorder and the current jetting effect are ruled out. The elongated geometry of the samples, together with silver paint contacts for current injection, encasing the whole ends of the wires, provide a homogeneous current injection through the entire bulk of the samples.[30,31] Theoretical simulations with similar sample geometries and even higher carrier mobility have shown[30] that magnetic fields much larger than 9 T are required for the onset of current jetting effects. To experimentally test for inhomogeneous current distribution in our wires, we use a special contact geometry on sample E (Extended Data Fig. 12). The voltage contacts on sample E were designed as point contacts, locally probing the voltage drop across two different edges of the sample. As in all other devices used in this study, the current injection contacts 1 and 6 cover the whole ends of the $(TaSe_4)_2I$ crystal. We measured simultaneously the potential difference $V_{i-j}$ between the two pairs of nearest-neighbor contacts along the current flow ($V_{2-3}$ and $V_{4-5}$) at $T = 80$ K, as a function of applied current $I$ and $B$ for the perpendicular ($I \perp B$) and longitudinal $I$ ($E \| B$) measurement configuration (Extended Data Fig. 12 (b)-(g) and (i)-(n)) The obtained curves are nearly identical and only display slight within the measurement error (up to 10 %). The agreement across the two pairs of contacts shows that distortions of the current path are minimal and that the current distribution is uniform within the sample. This implies that the observed positive longitudinal $\Delta dI/dV(B)$ is an intrinsic electronic effect. (ii) The absence of a $\Delta dI/dV(B)$ at perpendicular $B$, as well as the large temperatures at which the longitudinal $\Delta dI/dV(B)$ occurs, show that neither quantum interference effects of the condensate nor of the single quasiparticle excitations play a significant role. And (iii), the independence of $V_{th}$ on $B$ (Fig. 3 (a) and (e)) excludes any possible effects from a reduction of $V_{th}$.

**DATA AVAILABILITY STATEMENT**

All data generated or analyzed during this study are available within the paper and its Extended Data Files. Reasonable requests for further source data should be addressed to the corresponding author.

**ACKNOWLEDGEMENTS**

C.F. acknowledges the research grant DFG-RSF (NI616 22/1): Contribution of topological states to the thermoelectric properties of Weyl semimetals and SFB 1143. Z. W., and B. A. B. were supported by the Department of Energy Grant No. DE-SC0016239, the National Science Foundation EAGER Grant No. NOA-AWD1004957, Simons Investigator Grants No. ONR-N00014-14-1-0330 and No. NSF-MRSEC DMR- 1420541, the Packard Foundation, the Schmidt Fund for Innovative Research. Z.W. acknowledges support from the National Thousand-Young-Talents Program, the CAS Pioneer Hundred Talents Program, and the National Natural Science Foundation of China.

**AUTHOR INFORMATION**

**Contributions**

B.A.B., C. F. and J.G. conceived the experiment. N. K., Ch. S. and Y. Q. synthesized the single-crystal bulk samples. J.G., S.S.H and Cl.S. fabricated the electrical transport devices. J.G. carried out the transport measurements with the help of S.S.H. and Cl.S.. J.G. and C.S. analyzed the data. J.N., W.S., Z.J. and Y.S. calculated the band structure. B.B. and B.A.B. provided the theoretical background of the work. All authors contributed to the interpretation of the data and to the writing of the manuscript.

**Competing financial interest**

The authors declare no competing financial interests.

**Corresponding author**

* johannes.gooth@cpfs.mpg.de

**EXTENDED DATA FIGURES**

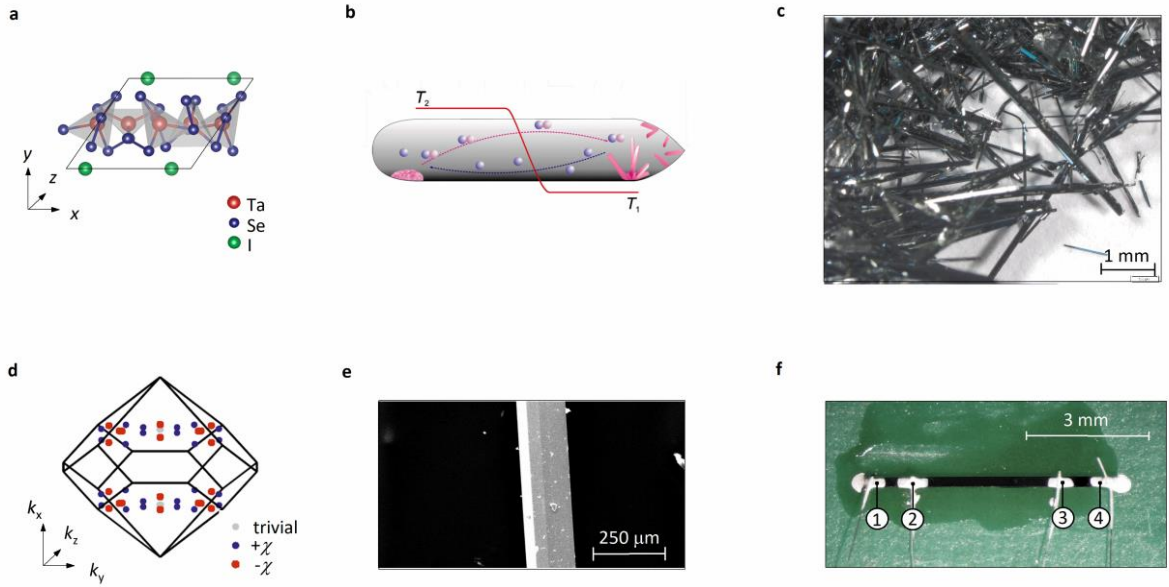

**Extended Data Figure 1 | (TaSe$_4$)$_2$I crystal structure, growth and device. a,** The quasi-one-dimensional crystal structure of (TaSe$_4$)$_2$I along the *x*-direction. **b,** Sketch of the growth principle. Single crystals of Ta$_2$Se$_8$I were obtained via a chemical vapor transport method using Ta, Se and I as starting materials and I$_2$ as carrier. A mixture with composition Ta$_2$Se$_8$I was prepared and sealed in an evacuated quartz tube. The ampoule was inserted into a furnace with a temperature gradient of $T_2 = 500$ to $T_1 = 400°C$ with the educts in the hot zone. After two weeks, needle-like crystals had grown in the cold zone. **c,** Optical micrograph of the growth substrate. **d**, Distribution of Weyl points in momentum (*k*)-space of opposite chirality $\pm\chi$. **e,** Scanning electron microscope image of a selected crystal. **f,** Typical device for electrical transport measurements. Contacts 1 and 4 are used to apply an electrical current. Contacts 2 and 3 are used to probe the corresponding voltage drop across the sample.

35**EXTENDED DATA FIGURES**

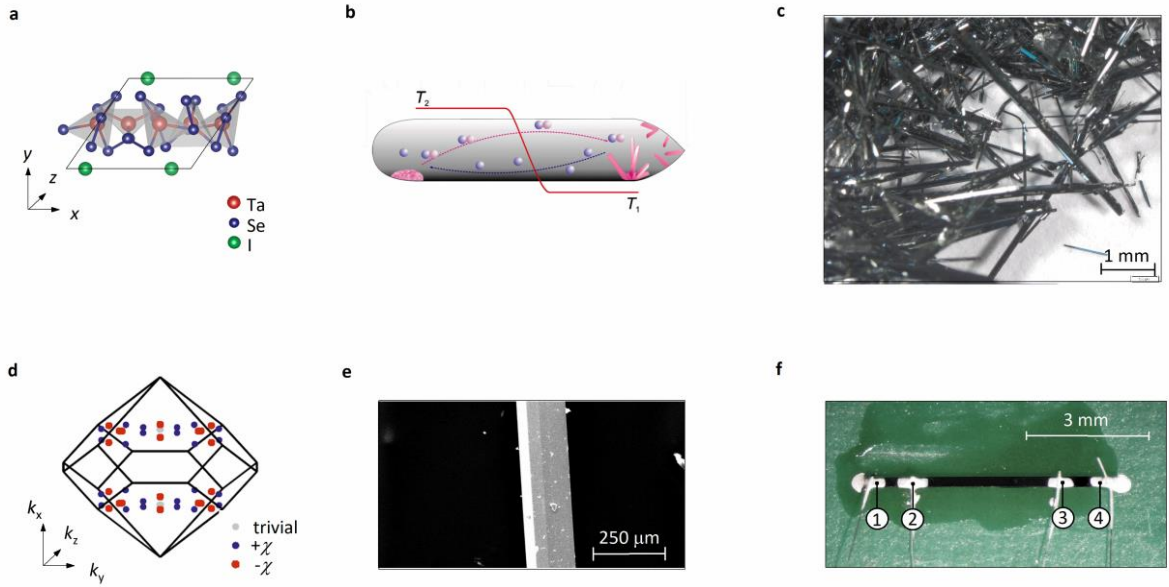

**Extended Data Figure 1 | (TaSe$_4$)$_2$I crystal structure, growth and device. a,** The quasi-one-dimensional crystal structure of (TaSe$_4$)$_2$I along the *x*-direction. **b,** Sketch of the growth principle. Single crystals of Ta$_2$Se$_8$I were obtained via a chemical vapor transport method using Ta, Se and I as starting materials and I$_2$ as carrier. A mixture with composition Ta$_2$Se$_8$I was prepared and sealed in an evacuated quartz tube. The ampoule was inserted into a furnace with a temperature gradient of $T_2 = 500$ to $T_1 = 400°C$ with the educts in the hot zone. After two weeks, needle-like crystals had grown in the cold zone. **c,** Optical micrograph of the growth substrate. **d**, Distribution of Weyl points in momentum (*k*)-space of opposite chirality $\pm\chi$. **e,** Scanning electron microscope image of a selected crystal. **f,** Typical device for electrical transport measurements. Contacts 1 and 4 are used to apply an electrical current. Contacts 2 and 3 are used to probe the corresponding voltage drop across the sample.

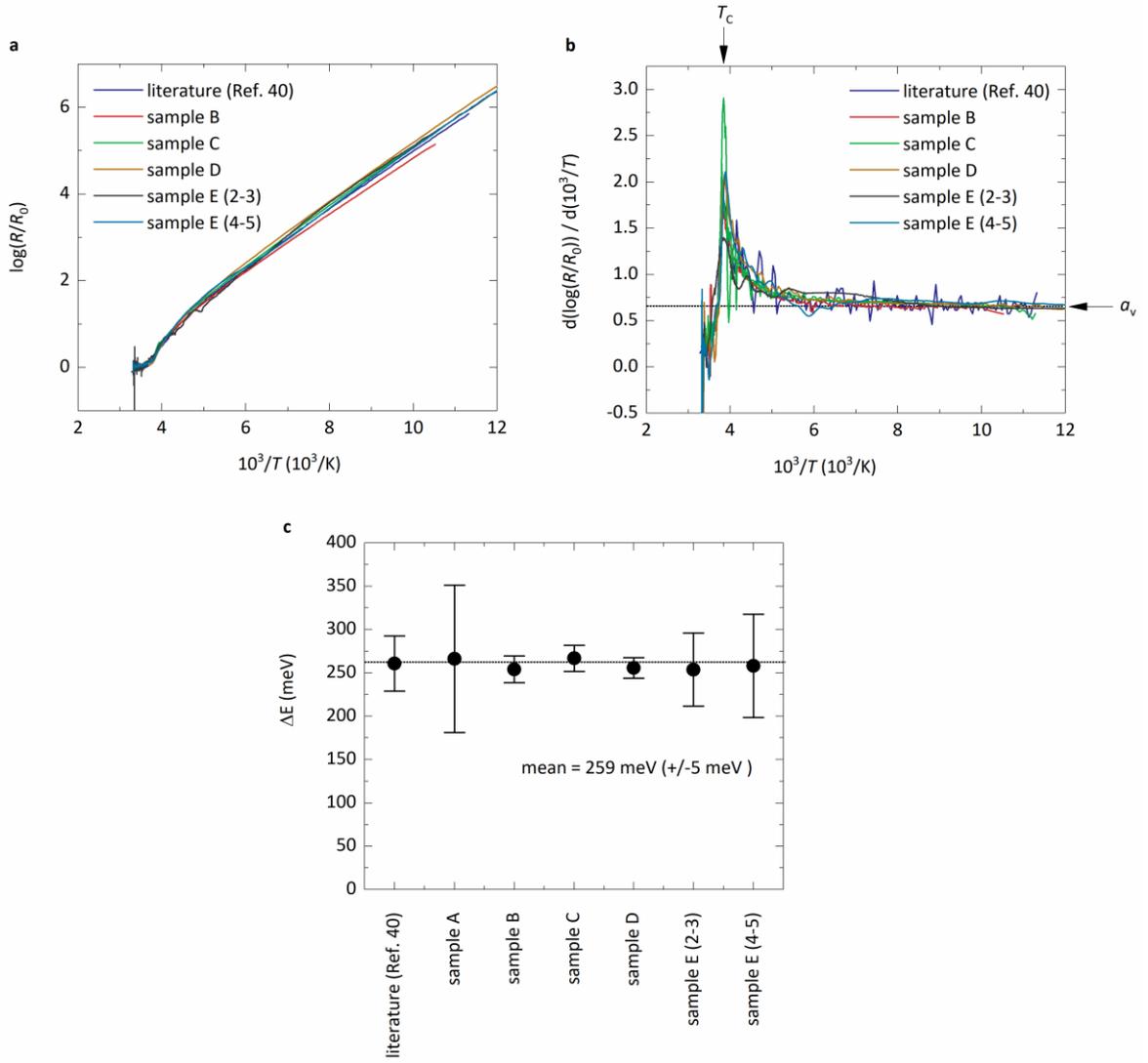

**Extended Data Figure 2 | Single-particle electrical resistance of samples B, C, D and E.** The numbers in the brackets mark the voltage contacts used on sample E, according to the sketches of the measurement configuration in Extended Data Fig. 12 and **a,** The electrical resistance $R$, normalized by its 300K-value $R_0$ and **b**, its logarithmic derivative as a function of $10^3/T$, where $T$ is the temperature in Kelvin. The derivative peaks at the CDW transition temperature $T_C = 263$ K and exhibits a mean-field form $R \sim \exp(\Delta E/k_B T)$ at below $T_C$, indicated by the $T$-independent value $a_v$ at low temperatures. **c,** Single-particle gaps of all (TaSe$_4$)$_2$I samples investigated. The single-particle gap $\Delta E$ is calculated f $a_v$. The error denotes the fit error. The dotted line displays the mean value of all sample A, B, C, D and E.

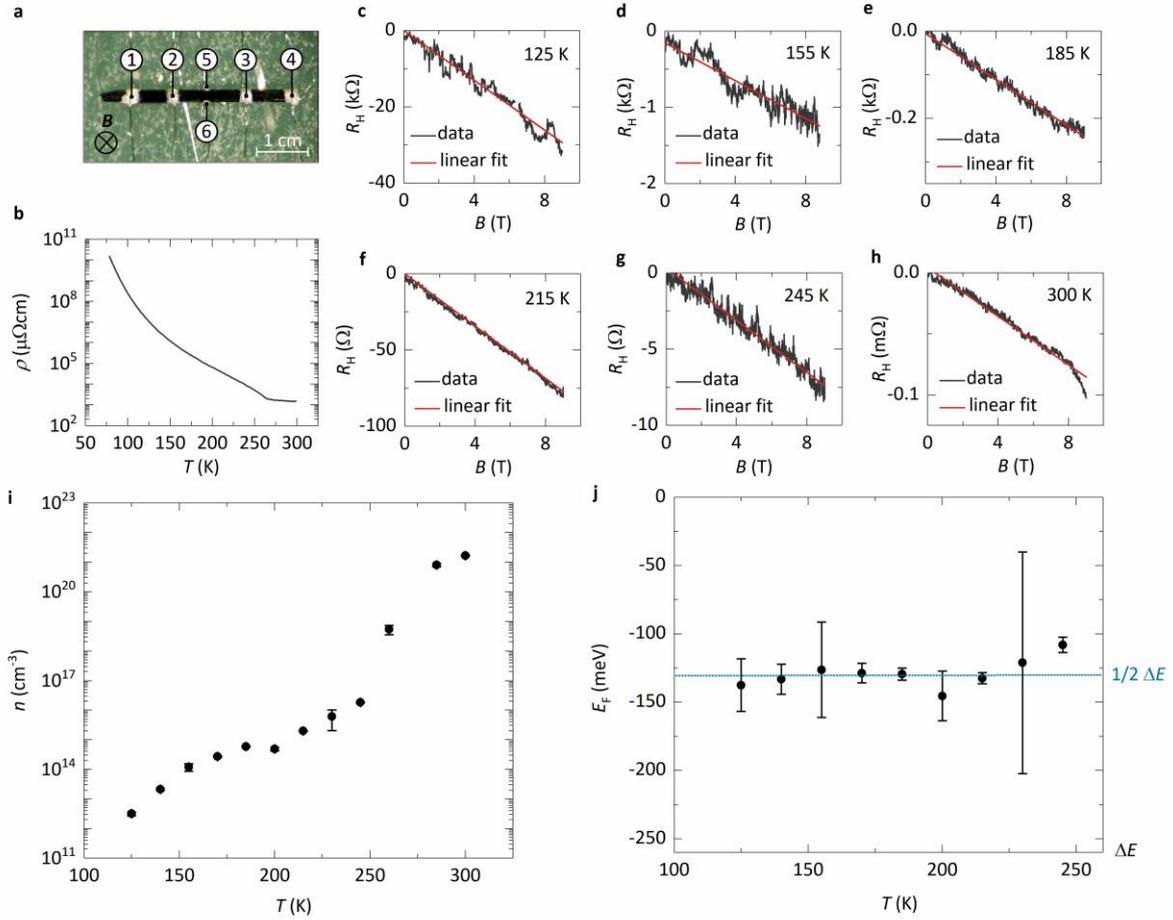

**Extended Data Fig. 3 | Single-particle Hall measurements and Fermi level position at selected temperatures of sample D. a,** Device for the Hall measurements. Contacts 1 and 4 are used for electrical current injection. Contacts 2 and 3 are used to probe the longitudinal voltage along the sample. Contacts 5 and 6 are used to measure the Hall voltage $V_{5,6}$ across the sample. The magnetic field $B$ is applied perpendicular to the measurement plane. **b,** Single-particle longitudinal resistivity $\rho$ versus temperature of sample D. **c,** Single particle Hall resistance $R_H = V_{5,6}/I$ at 125 K ($I$ = 1 μA), **d,** at 155 K ($I$ = 10 μA), **e,** at 185 K ($I$ = 10 μA), **f,** at 215 K ($I$ = 100 μA), **g,** at 245 K ($I$ = 100 μA) and **h,** at 300 K ($I$ = 2 μA). The black curves denote the measurement data and the red lines are linear fits to the data. The bias currents at each temperature are chosen low enough to probe only the single-particle transport, but high enough to resolve a Hall signal. Below 125 K, the longitudinal resistance of the sample became

too high to measure the Hall response. **i**, Carrier concentration $n$ as a function of temperature. $n$ is estimated from the slope $dR_H/dB$ of the linear fits to the magnetic field-dependent $R_H$ using the single charge carrier model $n = (dR_H/dB)^{-1}(e\,d)^{-1}$, where e is the elementary charge and $d$ = 300 μm the height of the sample. The error bars denotes the error of the linear fits. Estimated Fermi level position below $T_C$ with respect to the conduction band edge. The blue dotted line denotes the middle of the energy gap $\Delta E$ extracted from the single-particle electrical resistance in Extended Data Fig. 2.

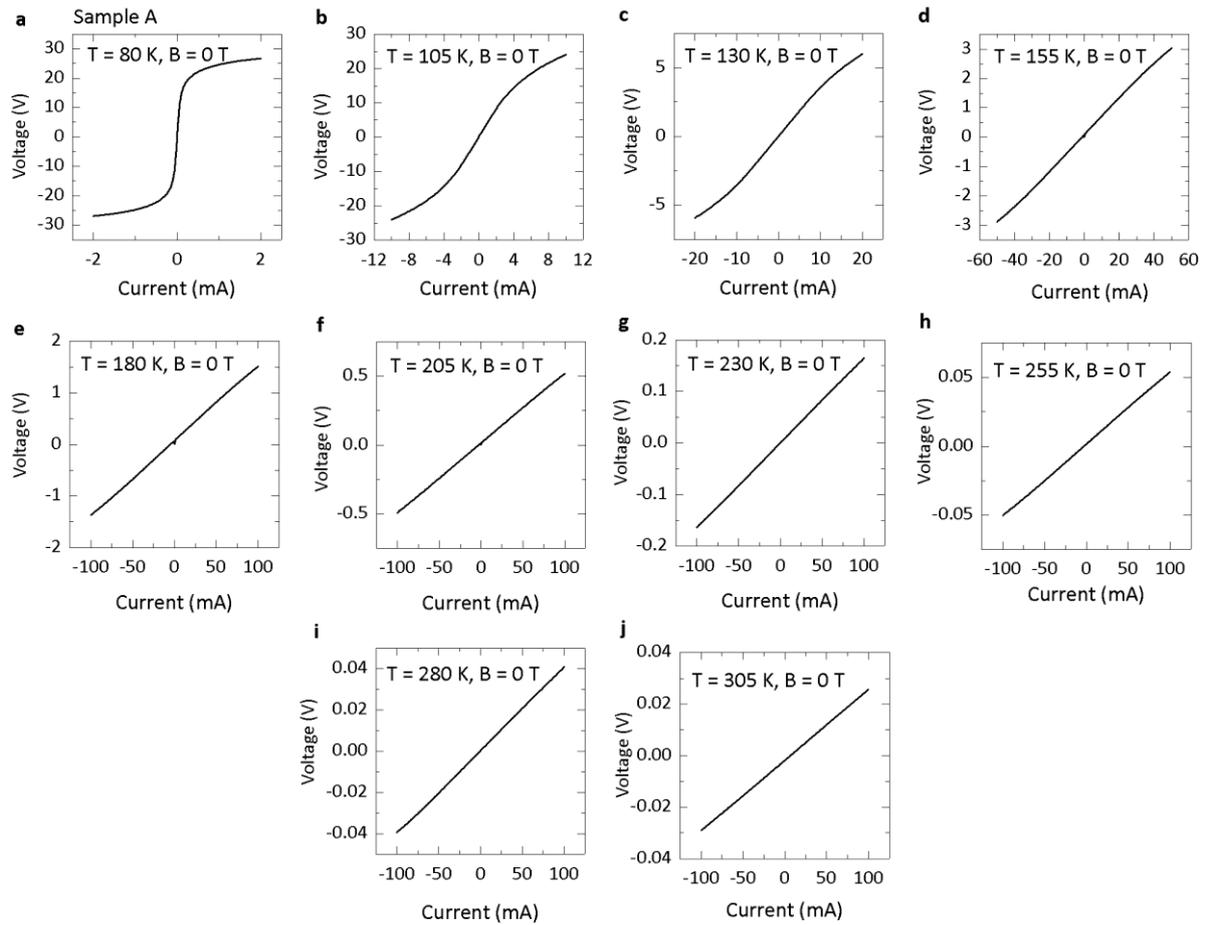

**Extended Data Figure 4 | *V-I* characteristics at selected temperatures of sample A at zero magnetic field. a,** *V-I* characteristics at 80 K; **b**, 105 K; **c**, 130 K; **d**, 155 K; **e** 180 K; **f**, 205 K; **g**, 230 K; **h**, 255 K; **i**, 280 K; and **j**, 305 K. This data is used to calculate the differential resistance d*V*/d*I* curves in Fig. 2 (d). At temperatures below 180 K, we start to observe non-linearity. These non-linearity become even more apparent in the d*V*/d*I*-curves shown in Fig. 2 (d), where a deviation from the linear behavior is already seen at 230 K.

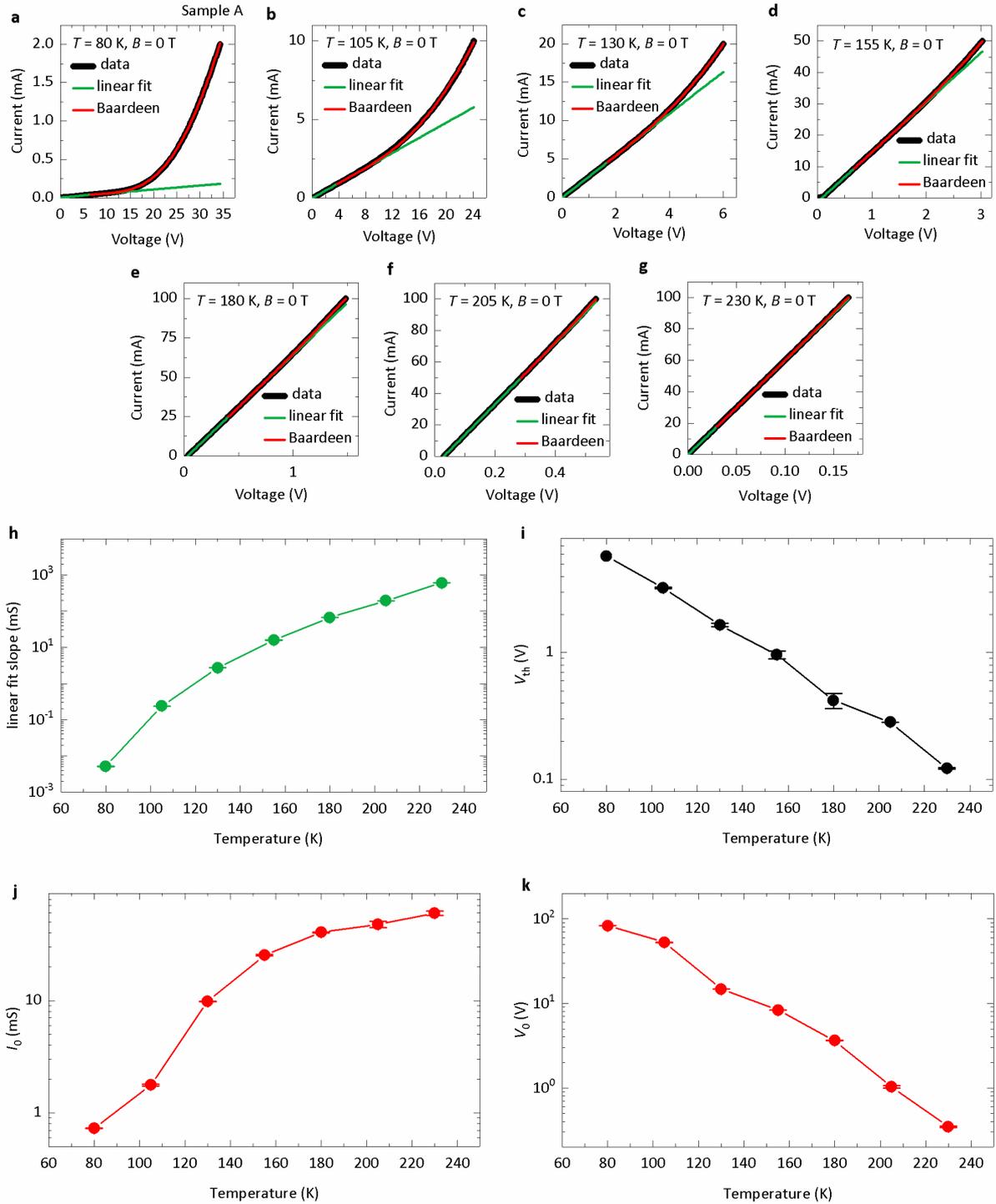

**Extended Data Figure 5 | Fitting the *V-I* characteristics at temperatures below $T_C$ of sample A at zero magnetic field. a,** *V-I* characteristics data (black line), a linear fit to the data around $V = 0$ V accounting for the single particle current contribution $I_{sp}(V) = mV$ (green line) and a fit of a model based on the Bardeen theory[45] $I(V) = I_{sp}(V) + I_{CDW}(V)$ (red line), where

$I_{CDW}(V) = I_0(V-V_T)\exp(-V_0/V)$ with constants $I_0$ and $V_0$ at 80 K; **b**, 105 K; **c**, 130 K; **d**, 155 K; **e** 180 K; **f**, 205 K and **g**, 230 K. Above the threshold voltage $V_T$, the measurements data is well represented by the Baardeen theory-fits. **h**, The slope *m* of the linear fits and **i**, the threshold voltage $V_{th}$ extracted from Fig. 2 have been employed to extract the fit parameters **j**, $I_0$ and **k**, $V_0$ as a function of temperature. The error bar denotes the error of the fits.

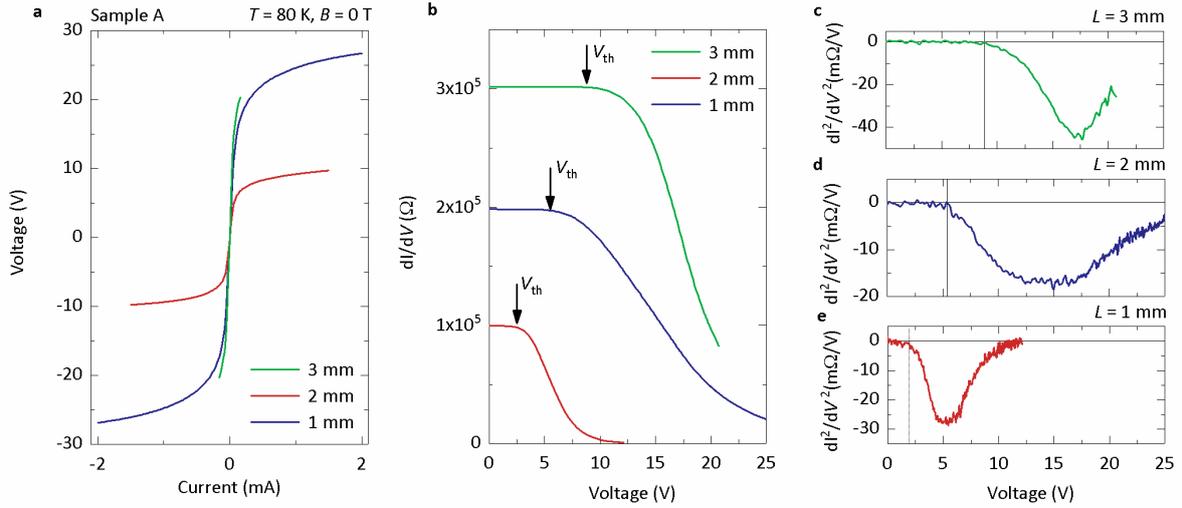

**Extended Data Figure 6 | Contact separation-dependence of the switching voltage of sample A. a,** $V$-$I$ characteristics and **b**, differential resistance $dV/dI$ at 80 K and zero magnetic field $B$ for various contact length $L$ on sample A. The threshold voltages $V_{th}$, shown in Fig. 2 (g) are estimated from the second derivatives $dV^2/dI^2$ displayed in **c**, **d** and **e**. We define $V_{th}$ as the voltage at which the last data point of the $dV^2/dI^2$ (marked by the vertical line) touches the zero-baseline before the global minimum upon enhancing $V$.

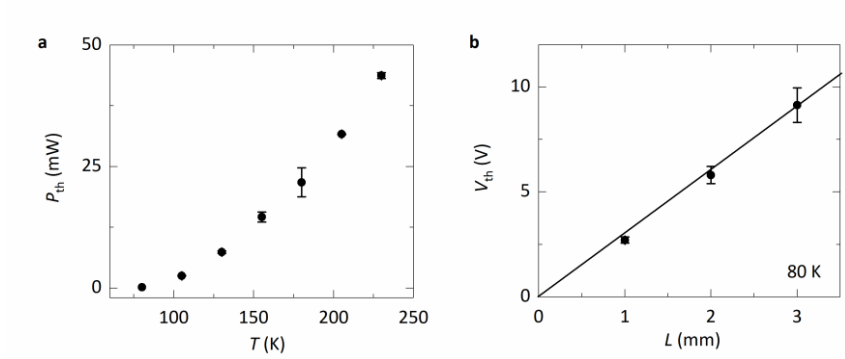

**Extended Data Figure 7 | Testing the origin of the non-linear IV curves. a**, Threshold electric field $E_{th} = V_{th}/L$ as a function of $T$ calculated from the $T$-dependent $V_{th}$ (Extended Data Fig. 5).The error bars denote the variation between $+V_{th}$ and $-V_{th}$ due to hysteresis effects. **b**, The increasing Joule heating power $P_{th} = V_{th}^2/[dV/dI(V_{th})]$ at the threshold voltage, as a function of increasing $T$ and **g**, the linear-dependence of $V_{th}$ on $L$ shown at 80 K (the line denotes a linear fit) demonstrate that the observed effects are intrinsic of the $(TaSe_4)_2I$ crystals.

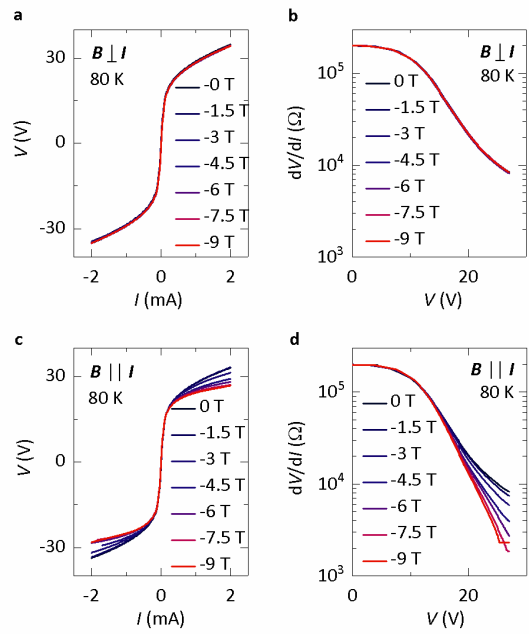

**Extended Data Figure 8 | Symmetry of the *V-I* characteristics in the magnetic field of sample A. a,** *V-I* characteristics and **b**, differential resistance d*V*/d*I* at 80 K for sample A in magnetic fields perpendicular to the applied current, but opposed to the magnetic field in Fig. 3 (a), (b), (e) and (f). **c,** *V-I* characteristics and **d**, differential resistance d*V*/d*I* at 80 K for sample A in magnetic fields parallel to the applied current. This data is used to calculate the *MR*, displayed in Fig. 3 (g). The temperature-variation of the magnetic field-dependent *V-I* characteristics is shown in Extended Data Fig. 9.

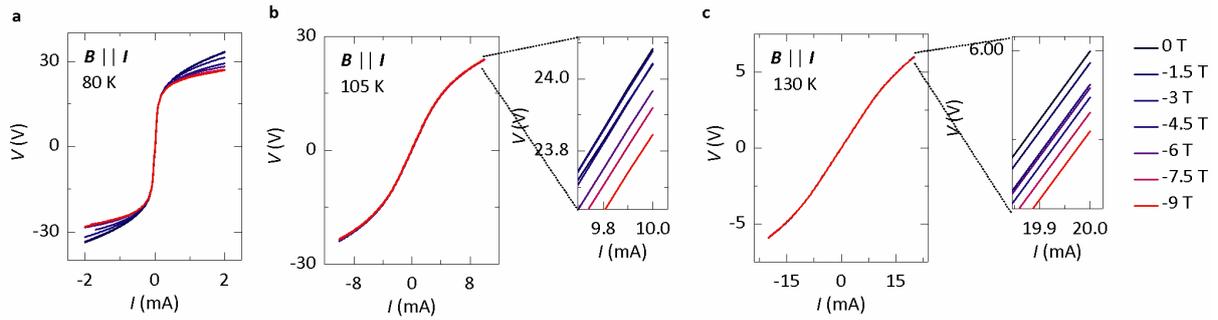

**Extended Data Figure 9 | Temperature-dependence of the magnetic field-variation of sample A for *I*||*B*. a**, *V-I* characteristic at 80 K **b**, 105 K and **c**, 130 K. At the highest bias currents applied, the voltage variation is around 30 % at 80 K, around 2 % at 105 K and around 0.5 % at 130 K.

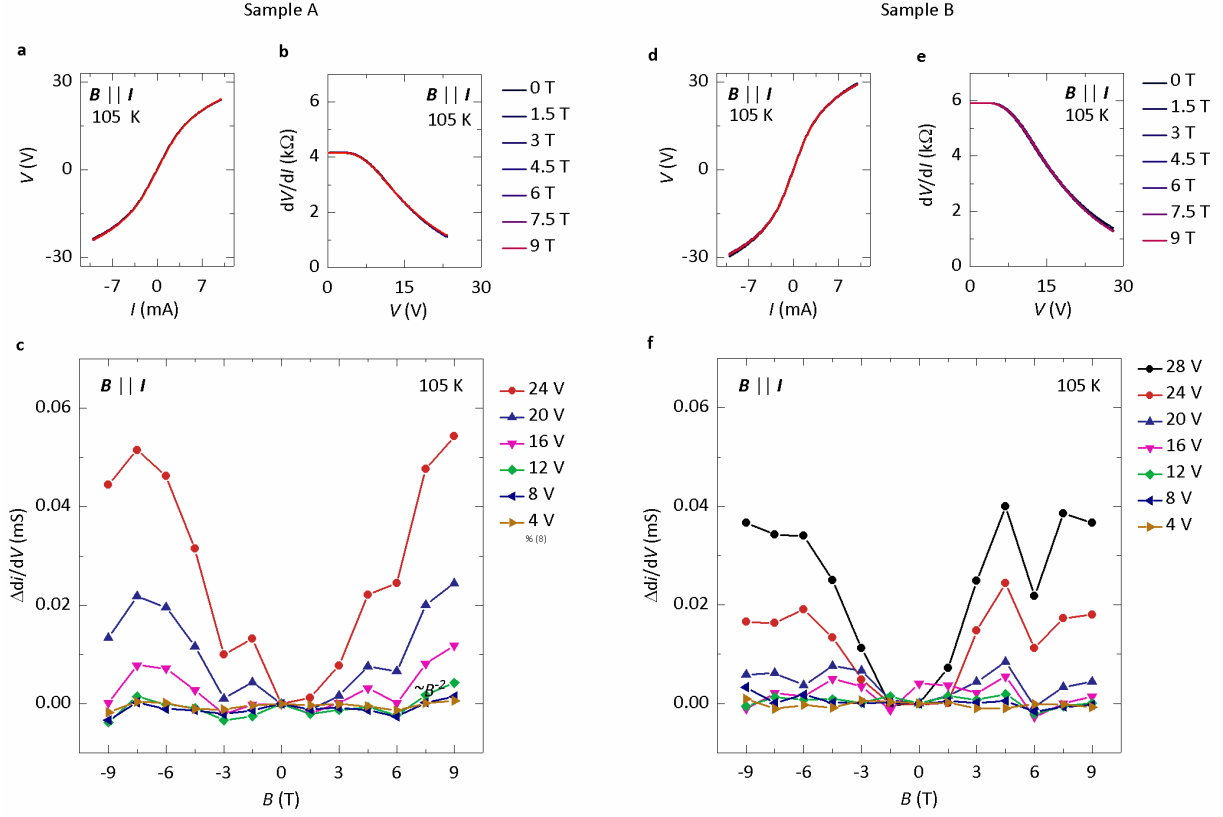

**Extended Data Figure 10 | Bias-dependent data of sample A and B at 105 K.** Both samples have similar contact separation, but sample B exhibits a smaller cross-section than sample A. **a**, *V-I* characteristic and **b**, d*V*/d*I*, at various values of the magnetic field *B*, applied perpendicular to the current direction (*I*⊥*B*) of sample A. **c**, *B*-dependence of the magneto-conductance Δd*I*/d*V*(*B*) = d*I*/d*V* (*B*) - d*V*/d*I* (0 T) at 105 K and at various *V* of sample A. **d**, *V-I* characteristic and **e**, d*V*/d*I*, at various values of the magnetic field *B*, applied perpendicular to the current direction (*I*⊥*B*) of sample B. **f**, *B*-dependence of the magneto-conductance Δd*I*/d*V*(*B*) = d*I*/d*V* (*B*) - d*V*/d*I* (0 T) at 105 K and at various *V* of sample B. The consistent observation of a positive longitudinal Δd*I*/d*V*(*B*) at high biases in both samples confirm our magnetic field-dependent experiments.

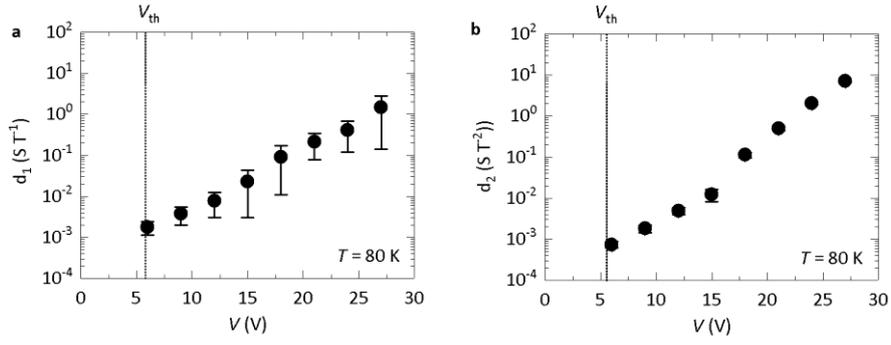

**Extended Data Figure 11 | Fit parameters of the** second-order polynomial fits $d_1 B_{//} + d_2 B_{//}^2$ to the experimental longitudinal $\Delta dI/dV(B_{//})$ shown in Fig. 3 (g).

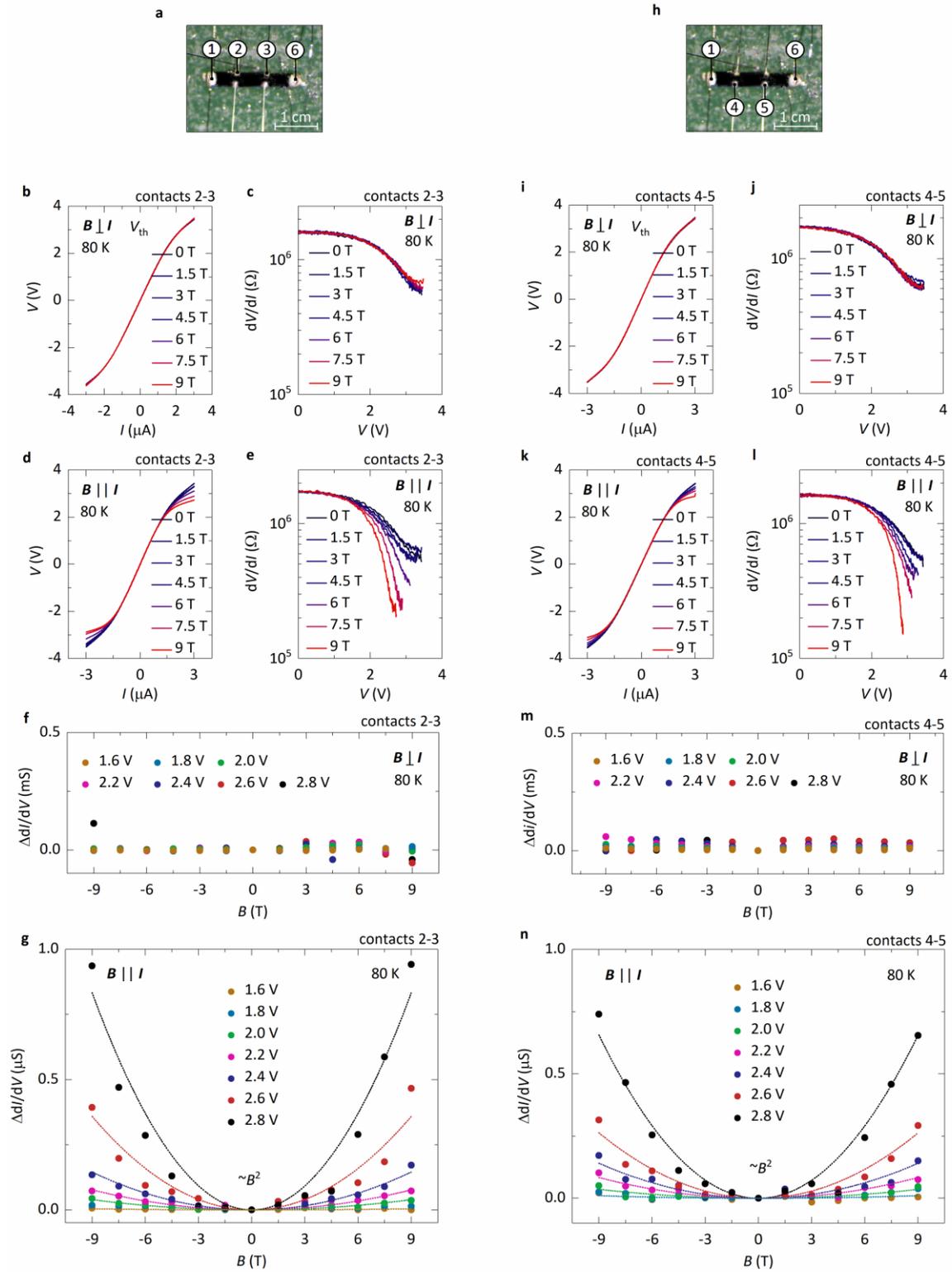

**Extended Data Figure 12 | Homogeneity-test of the current distribution on sample E.** The transport measurements are carried out in a magnetic field $B$. **a**, Sketch of the first measurement configuration: The electrical current is injected from contacts 1 and 6 and the voltage drop is measured between contacts 2 and 3. **b**, $V$-$I$ characteristic and **c**, d$V$/d$I$, at 80 K and at various values of the magnetic field $B$, applied perpendicular to the current direction ($I \perp B$) and **d**, **e**, applied parallel to the current direction ($I||B$) for the first measurement configuration . **f**, $B$-dependent variation of the differential conductance $\Delta \mathrm{d}I/\mathrm{d}V(B) = \mathrm{d}I/\mathrm{d}V(B) - \mathrm{d}V/\mathrm{d}I(0\,\mathrm{T})$ at 80 K and at various $V$ for $I \perp B$ and **g,** for $I||B$. **h**, Sketch of the second measurement configuration: The electrical current is injected from contacts 1 and 6 and the voltage drop is measured between contacts 4 and 5. **i**, $V$-$I$ characteristic and **j**, d$V$/d$I$, at 80 K and at various values of the magnetic field for ($I \perp B$) and **k**, **l**, ($I||B$) for the second measurement configuration . **m**, $B$-dependent variation of the differential conductance $\Delta \mathrm{d}I/\mathrm{d}V(B)$ at 80 K and at various $V$ for $I \perp B$ and **n,** for $I||B$ for the second measurement configuration. The consistent observations is both measurement configurations with point contacts indicates a homogeneous distribution of the electrical current and confirms our magnetic field-dependent experiments.